\documentclass[prd,preprint,floatfix,eqsecnum,nofootinbib,12pt]{revtex4}

\usepackage{amssymb,amsmath,amsthm,graphicx,ulem}
\pdfoutput=1
\usepackage{graphicx,subfigure}
\usepackage{ulem}
\usepackage{epsfig}
\usepackage{amsmath}
\usepackage{amsfonts}
\usepackage{amssymb}
\usepackage[usenames]{color}
\usepackage[letterpaper,left=2.cm,right=2.cm,top=2.5cm,bottom=2.5cm]{geometry}
\usepackage[T1]{fontenc}

\newcommand{\beq}{\begin{equation}}
\newcommand{\eeq}{\end{equation}}
\newcommand{\be}{\begin{equation}}
\newcommand{\ee}{\end{equation}}
\newcommand{\beqa}{\begin{eqnarray}}
\newcommand{\eeqa}{\end{eqnarray}}
\newcommand{\beqar}{\begin{eqnarray*}}
\newcommand{\eeqar}{\end{eqnarray*}}
\newcommand{\bea}{\begin{eqnarray}}
\newcommand{\eea}{\end{eqnarray}}

\newcommand{\lp}{\left(}
\newcommand{\rp}{\right)}

\def\th{{\tilde h}}

\def\be{\begin{equation}}
\def\ee{\end{equation}}
\def\vx{{\vec x}}
\def\rx{\bf}

\newcommand{\nn}\nonumber

\begin{document}

\title{\textbf{Holographic Tunneling Wave Function}}
\author{Gabriele Conti\footnote{gabriele.conti@fys.kuleuven.be}, Thomas Hertog\footnote{thomas.hertog@fys.kuleuven.be}, Ellen van der Woerd\footnote{ellen@itf.fys.kuleuven.be}}
\affiliation{Institute for Theoretical Physics, University of Leuven, 3001 Leuven, Belgium}

\date{\today}

\bibliographystyle{unsrt}

\begin{abstract}

The Hartle-Hawking wave function in cosmology can be viewed as a decaying wave function with anti-de Sitter (AdS) boundary conditions. We show that the growing wave function in AdS familiar from Euclidean AdS/CFT is equivalent, semiclassically and up to surface terms, to the tunneling wave function in cosmology. The cosmological measure in the tunneling state is given by the partition function of certain relevant deformations of CFTs on a locally AdS boundary. We compute the partition function of finite constant mass deformations of the $O(N)$ vector model on the round three sphere and show this qualitatively reproduces the behaviour of the tunneling wave function in Einstein gravity coupled to a positive cosmological constant and a massive scalar. We find the amplitudes of inhomogeneities are not damped in the holographic tunneling state.

\end{abstract}

\maketitle

\section{Introduction}

The dS/CFT proposal \cite{Balasubramanian2001,Strominger2001,Maldacena2002,Witten2001} conjectures that `the' wave function of the universe with asymptotic de Sitter (dS) boundary conditions is given in terms of the partition function of a Euclidean CFT deformed by various operators. But which wave function of the universe does the CFT select? Alternatively dS/CFT may be able to accommodate different wave functions. Some evidence for this comes from the explicit higher spin version of dS/CFT \cite{Anninos2011} where the partition functions appear to exhibit certain features reminiscent of the Hartle-Hawking wave function in Einstein gravity \cite{Anninos2012}, whereas other properties have a natural interpretation in the tunneling state \cite{Conti2014}.

Some insight in this question comes from the holographic form of the semiclassical Hartle-Hawking wave function derived in \cite{Hertog2011}, building on the wave function approach to dS/CFT pioneered in \cite{Maldacena2002} and further explored in \cite{Garriga2008,Harlow2011,Maldacena2011,Castro2012,Banerjee2013,Anninos2012,Anninos2013,Anninos2014,Hartle2012a,Hartle2012b}. The holographic Hartle-Hawking wave function reveals an intricate connection between dS/CFT and Euclidean AdS/CFT. This arises because all complex saddle points associated with asymptotically locally dS universes have a representation in which their interior geometry consists of a Euclidean, locally AdS space that makes a smooth transition to a Lorentzian, asymptotically dS universe. Moreover the finite part of the wave function that defines the Hartle-Hawking measure is fully specified by the AdS regime of the saddle points. The transition region between AdS and dS merely compensates for the volume terms in the AdS action and accounts for the phases that explain the classical behavior of the final configuration \cite{Hertog2011}. 
This means one can use Euclidean AdS/CFT to write the semiclassical Hartle-Hawking measure on the space of classical, inflationary, asymptotically dS histories in terms of the partition function of a dual CFT defined on the boundary and deformed by certain (relevant) deformations\footnote{The asymptotic dS structure acts as a final condition that sets all sources corresponding to irrelevant deformations to zero. We discuss this more in Section \ref{holo}.}. In this description, the sources of the partition function correspond to the argument of the asymptotic wave function.  

The potential in the Wheeler-DeWitt (WDW) equation in an asymptotically AdS context is everywhere positive. Generic solutions $\Psi$ thus have a growing and a decaying branch in the large volume regime.
A closer look at the behavior of the Hartle-Hawking wave function in its AdS domain shows it corresponds to a very special, decaying wave function in AdS \cite{Hartle2012a,Hartle2012b}. This is closely connected to the fact that the holographic Hartle-Hawking measure involves the inverse of the AdS/CFT dual partition function \cite{Hertog2011}.

Here we investigate whether there are reasonable quantum states in cosmology corresponding to the growing branch of AdS wave functions. In a sense one might argue such states have a cleaner holographic interpretation, because the dual theory in AdS/CFT most directly computes the growing branch of bulk wave functions only. Indeed, the local surface terms one usually adds to the bulk action to extract information essentially eliminate the decaying branch. 

We find that Vilenkin's tunneling wave function \cite{Vilenkin1987,Vilenkin1986}, in the semiclassical approximation and up to local surface terms, is equivalent to the growing branch of the most natural wave function in AdS, in which the amplitudes of boundary configurations are specified by the partition function of a dual CFT with certain finite deformations. To show this we again exploit the complex structure of the bulk saddle points. This allows us to establish a relation between the tunneling wave function with asymptotic dS boundary conditions and an AdS wave function which we interpret in the context of AdS/CFT. As a first test of our proposal for a holographic tunneling state we compute the partition function of the $O(N)$ vector model on the round three sphere as a function of a homogeneous, finite mass deformation. We do this both for the critical and for the minimal model. We find the partition functions in both cases qualitatively agree with the behaviour of the minisuperspace tunneling wave function in Einstein gravity coupled to a positive cosmological constant and a massive scalar field, and we interpret this as evidence in favour of our proposal.

As solutions to the WDW equation the tunneling and Hartle-Hawking wave functions obey different boundary conditions. As a consequence they specify a different cosmological measure, which translates into different predictions for what we should expect to observe\footnote{See e.g. \cite{Hertog2013} for a sharp illustration of this.}. A controversial feature of the tunneling wave function has been whether it actually predicts that the amplitude of inhomogeneous fluctuations is suppressed, as required by observation. In order for this to be the case the fluctuation action must enter in the wave function with a different sign than the action for the background histories. This sign change has been motivated on the basis of regularity of the wave function \cite{Vilenkin1987, Vachaspati1988, Garriga1997}. We revisit this point in the Discussion below and conclude there is no evidence for this from a holographic viewpoint.
 
%%%%%%%%%%%%%%%%%%
\section{The Tunneling Wave Function}
%%%%%%%%%%%%%%%%%%%

We consider Einstein gravity coupled to a positive cosmological constant $\Lambda$ and a scalar field moving in a positive potential $V$. A quantum state of the universe in this model is given by a wave function $\Psi$ on the superspace of all three-metrics $h_{ij}(\vec{x})$ and field configurations $\chi(\vec{x})$ on a closed spacelike surface $\Sigma$,
\be
\Psi[h(\vec{x}),\chi(\vec{x})].
\ee
All wave functions must satisfy an operator implementation of the classical constraints. These include the Wheeler-DeWitt (WDW) equation ${\cal H} \Psi=0$. To solve the WDW equation one must specify boundary conditions on $\Psi$. A choice of boundary conditions specifies a quantum state of the universe, which together with the dynamics provides a predictive framework of cosmology.

\begin{figure}
\begin{center}
\includegraphics[width=8cm]{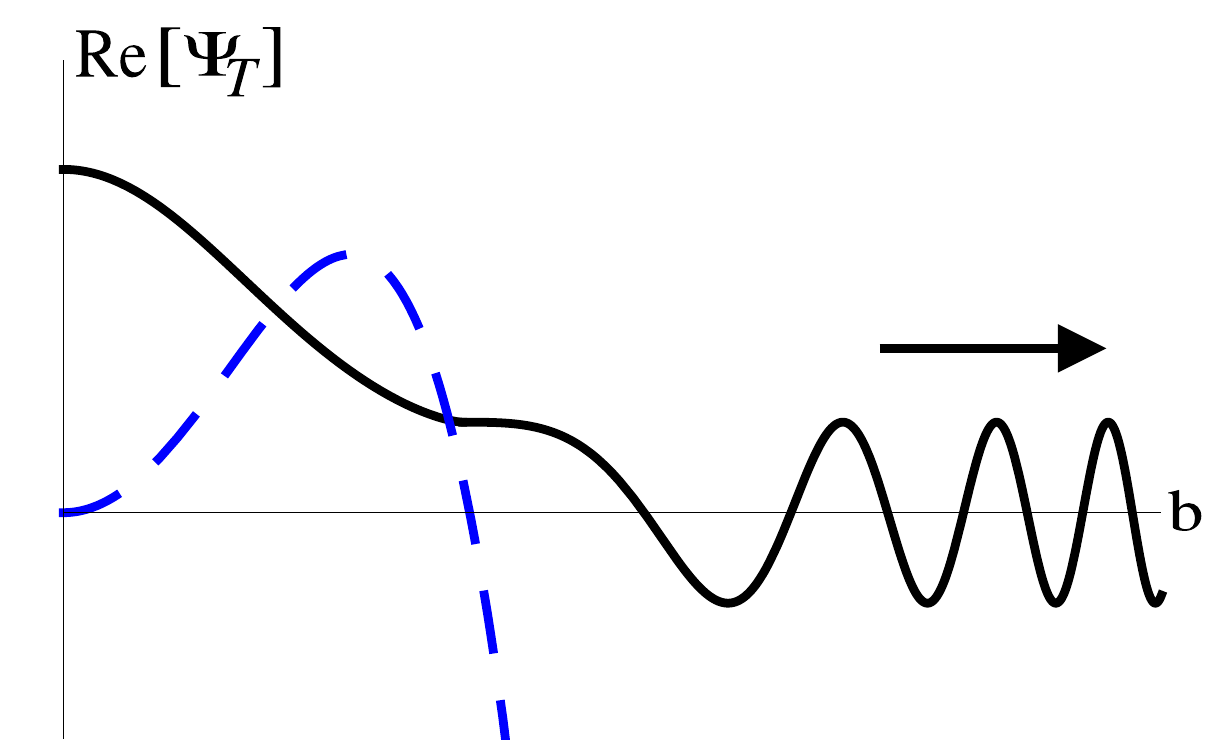} \hfill
\includegraphics[width=8cm]{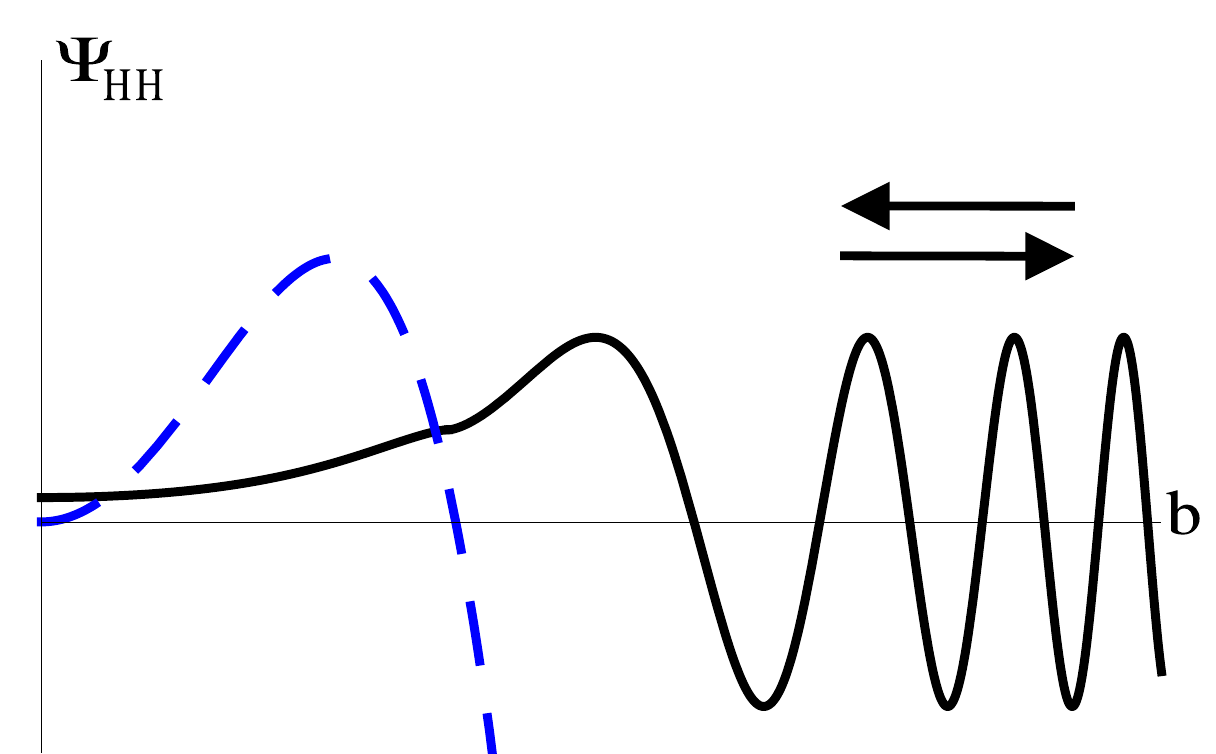} 
\end{center}
\caption{Qualitative behaviour of the tunneling wave function $\Psi_T$ (left) and the Hartle-Hawking wave function $\Psi_{HH}$ (right) in minisuperspace, as a function of the scale factor $b$. The wave functions obey different boundary conditions and therefore predict different cosmological measures. The blue dashed line shows a slice of the superpotential $U$ in the presence of a positive scalar effective potential.}
\label{pot}
\end{figure}

In the tunneling approach to quantum cosmology \cite{Vilenkin1986,Vilenkin1987}, the boundary condition on the wave function is that $\Psi_T$ in the large three-volume regime of superspace should include only outgoing waves, describing expanding universes. Physically this boundary condition implements the idea that our expanding classical universe originated in a quantum tunneling event. The motivation for this comes from an analysis in homogeneous isotropic minisuperspace where the WDW equation has a potential $U(b)$ of the form (cf. Fig. \ref{pot})
\be
U(b) = b^2 -b^4 V_\Lambda(\chi),
\ee
with $b$ the scale factor and $V_\Lambda(\chi) \equiv \Lambda/3 + V(\chi)$. The WDW equation therefore has growing and decaying solutions in a classically forbidden region $0<b<b_c$ under the barrier, where $b_c = 1/\sqrt{V_\Lambda(\chi)}$, but it has outgoing and ingoing wave solutions - corresponding to expanding and contracting universes - for $b \geq b_c$. One can thus envision a tunneling process from $b=0$, or `nothing', to a closed universe of radius $b_c$ which then subsequently expands. This selects the outgoing solution at large scale factor. In the semiclassical approximation and at large volume the resulting wave function will then oscillate rapidly and be of the form
\be
\label{formT}
\Psi_T[b,\chi] = A [b,\chi] \exp\left(iS[b,\chi]\right)
\ee
where $S$ is large and negative. The tunneling probabilities ${\cal P} = A^2$ are approximately given by \cite{Vilenkin1987}
\be\label{nuc}
{\cal P} \sim \exp\left( -2\int_0^{b_c} db \sqrt{U(b)}\right) \approx \exp\left(\frac{-3\pi}{V_{\Lambda}(\phi_0)}\right),
\ee
where $\phi_0$ is the absolute value of the scalar at the $b=0$ boundary (cf. Section \ref{homo}).
Once a universe nucleates it evolves classically and expands in an inflationary manner. The boundary condition that the wave function should include only outgoing waves in the large volume limit implies it is a linear combination of the growing and decaying solutions in the classically forbidden region\footnote{A more formal definition of the tunneling wave function in terms of a Lorentzian path integral was put forward in \cite{Vilenkin1994}. However this can only be evaluated in the semiclassical limit we discuss here.}. However the nucleation probabilities \eqref{nuc} are specified by the decaying solution under the barrier \cite{Vilenkin1987}. 

%%%%%%%%%%%%%%%%
%%%%%%%%%%%%%%%%

This is in sharp contrast with the Hartle-Hawking boundary conditions which select the growing solution under the barrier. This yields a real linear combination of ingoing and outgoing waves in the large volume region, describing a time-symmetric ensemble of contracting and expanding universes. Relative probabilities in the Hartle-Hawking state are specified by the amplitude of the growing solution when it emerges from the classically forbidden region. As a result both wave functions differ in their predictions: the tunneling wave function \eqref{nuc} favours universes in which the scalar field starts high up its potential, leading to a long period of inflation, whereas the Hartle-Hawking wave function favours histories with a low amount of inflation \cite{Hartle1983}.

In the semiclassical approximation both wave functions can be evaluated using saddle points. These are complex solutions of the Einstein equation, discussed in more detail below, which interpolate between `nothing' at the South Pole of an approximate four-sphere of radius $\sim 1/\sqrt{V(\phi_0)}$ to a Lorentzian, classical, inflationary universe. The saddle points specifying $\Psi_T$ are the same as those defining the semiclassical Hartle-Hawking wave function. However they are weighted differently in both wave functions, because they behave very differently under the barrier. The Hartle-Hawking probabilities are of the form $\vert \Psi_{HH} \vert^2 \sim \exp(-2I_E)$, where $I_E \approx -3\pi/|V_{\Lambda}(\phi_0)|$ is the real part of the Euclidean saddle point action. By contrast it follows from \eqref{nuc} that the semiclassical tunneling probabilities are of the form $\vert \Psi_T \vert^2 =A^2 \sim \exp(+2I_E)$. 

It is subtle, however, to implement the tunneling boundary condition in the full superspace. A naive extension of the above framework to include perturbations around homogeneous isotropic configurations does not yield a normalizable probability distribution for perturbations. For this reason it has been argued \cite{Vilenkin1987, Vachaspati1988, Garriga1997} that the probabilities for perturbations in the tunneling state are given by the growing branch of the wave function under the barrier, as in the Hartle-Hawking state.  One motivation for our work is to revisit this point from a holographic perspective. We return to this in the Discussion.

%%%%%%%%%%%%%%%%%
\section{Representations of Complex Saddle Points}
%%%%%%%%%%%%%%%%%%

We briefly review the saddle point geometries defining the semiclassical tunneling wave function. The Lorentzian action $S_L$ of our model is given by the sum of the Einstein-Hilbert action\footnote{We work in Planck units where $\hbar = c= G=1$.}
\be \label{lact}
S_{L}^{(EH)} = \frac{1}{16 \pi} \int_{\mathcal{M}} d^4x \sqrt{-g}   \left( R-2\Lambda\right) + \frac{1}{8\pi}\int_{\Sigma} d^3x \sqrt{h} K
\end{equation}
and the matter action\footnote{The normalization of the scalar field $\phi$ has been chosen to simplify subsequent equations.}
\begin{equation}
S_{L}^{(m)} = \frac{3}{4 \pi} \int_{\mathcal{M}} d^4x \sqrt{-g}  \left[-\frac{1}{2}\lp \nabla \phi\rp ^2- V(\phi)\right].
\end{equation}

We write the line element of a closed three-geometry as
\begin{equation}
d\Sigma^2 = b^2 \tilde h_{ij}(\vec{x})dx^idx^j,
\label{bdmetric}
\end{equation}
where $b$ is an overall scale factor and we take $\tilde{h}_{ij}(\vec{x})$ to have unit volume in the large scale factor limit. Superspace is therefore spanned by $b$ and $\tilde{h}_{ij}(\vec{x})$, and the boundary configuration $\chi (\vec{x})$ of the scalar field $\phi$. Thus $\Psi=\Psi(b,\tilde{h}_{ij},\chi)$. 

The compact saddle point geometries that specify the semiclassical approximation to the wave function are of the form
\beq \label{sad}
ds^2=-N^2(\lambda)d\lambda ^2+g_{ij}(\lambda,\vec{x})dx^idx^j,
\eeq
where $\{\lambda,x^i\}$ are four real coordinates on the real manifold $\mathcal{M}$. We take $\lambda =0$ to locate the South Pole (SP) of the saddle point, where the geometry caps-off, and $\lambda=1$ to locate the boundary $\Sigma$ of $\mathcal{M}$. Regularity at the SP together with the boundary condition that the geometry and field matches the real boundary configuration $(b,\tilde{h}_{ij},\chi)$ on $\Sigma$ mean that the saddle points are generally complex solutions of the Einstein equation.

The Einstein equation can be solved for $\{g_{ij}(\lambda,\vec{x}), \phi(\lambda,\vec{x})\}$ for any complex $N(\lambda)$ that is specified. Different choices of $N(\lambda)$ yield different geometric representations of the same saddle point. If we define the complex variable 
\begin{equation}
\tau(\lambda) \equiv \int_0^{\lambda} d \lambda' N(\lambda') + \tau_0,
\end{equation}
then different choices of $N(\lambda)$ correspond to different contours in the complex $\tau$-plane. Contours begin at the SP at $\lambda =0$, with $\tau(0)\equiv \tau_0$, and end at the boundary $\lambda=1$, with $\tau(1)\equiv \upsilon$. Each contour that connects $\tau_0$ to $\upsilon$ yields a different representation of the same complex saddle point. This freedom in the choice of contour gives physical meaning to a process of analytic continuation --- not of the Lorentzian histories themselves --- but of the saddle points that define their probabilities.   

In the next section we recall that this can be used to identify two different useful representations of all saddle points corresponding to asymptotically locally de Sitter, classical, Lorentzian histories. In one representation (dS) the interior saddle point geometry behaves as if the cosmological constant and the scalar potential were positive. In the other (AdS) the Euclidean part of the interior geometry behaves as if these quantities were negative, and specifies an asymptotically locally AdS space. Asymptotically Lorentzian de Sitter universes and Euclidean anti-de Sitter spaces are thereby connected in the semiclassical wave function. 

The action of a saddle point is an integral of its complex geometry and fields, which includes an integral over complex time $\tau$. Different contours for this time integral each yield the same amplitude of the boundary configuration the saddle point corresponds to. This provides the basis for the holographic form of $\Psi_T$ as we now discuss.

%%%%%%%%%%%%%%%%%%%%%%%
\section{Homogeneous Minisuperspace}
%%%%%%%%%%%%%%%%%%%%%%%%
\label{homo}

We are interested in the tunneling wave function at large volume. In this regime the semiclassical approximation to the WDW equation holds and implies an asymptotic expansion of the wave function which is essentially equivalent to the asymptotic expansion of solutions to the Einstein equation with asymptotic dS boundary conditions \cite{Hartle2012a,Hartle2012b}. We can therefore work directly with the asymptotic saddle point equations to study the asymptotic structure of $\Psi_T$.

We begin by considering ${\cal O}(4)$ invariant saddle points for which $\tilde h_{ij}(\vec{x})$ is fixed to be the metric of a round unit three sphere $\gamma_{ij}$. Homogeneous and isotropic minisuperspace is thus spanned by the scale factor $b$ and the homogeneous value of the scalar field $\chi$, and $\Psi=\Psi(b,\chi)$. 

The line element of the four-geometries that contribute to the minisuperspace wave function can be written as
\begin{equation}
ds^2 = -d\tau^2 + a^2(\tau) \gamma_{ij} dx^i dx^j
\label{himet}
\end{equation}
and the saddle point equations implied by the action \eqref{lact} become
\begin{align}
\dot{a}^2 + 1 - H^2 a^2 - a^2 \dot{\phi}^2 - 2 a^2 V &= 0, \\
\ddot{\phi} + 3 \frac{\dot{a}}{a} \dot{\phi} + V_{,\phi} & = 0,
\end{align}
where $\dot a \equiv da/d\tau$ and $H^2 \equiv \Lambda/3$. 
Solutions specify functions $a(\tau)$ and $\phi(\tau)$ in the complex $\tau$-plane. A contour $C(\tau_0,\upsilon)$  representing a saddle point connects the SP at  $\tau_0$ to a point $\upsilon$ where $a(\upsilon)$ and $\phi(\upsilon)$ take the real values $b$ and $\chi$ respectively. For any such contour the on-shell action is given by  
\begin{equation}
S_L = - \frac{3\pi}{2}\int_{{C}(\tau_0,\upsilon)} d\tau a \left[ a^2\left(H^2+ 2V(\phi)\right)-1\right].
\label{eucact_alt}
\end{equation}

In terms of the variable $u \equiv e^{-H \tau} \equiv e^{-H(y+iz)}$ the large volume regime corresponds to the large $y$ limit. The asymptotic expansions in powers of $u$ of the scale factor and field are given by
\begin{align}
a(u) &= \frac{c}{u} \left[1+\frac{u^2}{4c^2 H^2} -\frac{3}{4}\alpha^2 u^{2\lambda_{-}} +\cdots - \frac{2m^2\alpha\beta}{3 H^2} u^3 + \cdots \right], \label{aa_series} \\
\phi(u) &= u^{\lambda _-}( \alpha + \alpha_1 u + \cdots )   +  u^{\lambda _+}(\beta + \beta_1 u  + \cdots),
\label{pphia}
\end{align}
where we have assumed that when the scale factor is large and the field is small, the scalar field potential is quadratic, i.e. $V(\phi) =(1/2)m^2\phi^2 + {\cal O}(\phi^4) + \cdots$, and $\lambda _\pm \equiv \frac{3}{2} \pm \sqrt{\frac{9}{4} - \frac{ m^2}{H^2}}$.

The complex asymptotic solutions are locally determined in terms of the argument of the wave function, i.e. the `boundary values' $c^2\gamma_{ij}$ and $\alpha$, up to the $u^3$ term in \eqref{aa_series} and to order $u^{\lambda_{+}^{\ }}$ in \eqref{pphia}. The further coefficients in the expansions encode information about the detailed physics of the matter (the shape of the potential), the interior of the saddle point geometry and the boundary condition of regularity at the SP. We now use this complex asymptotic structure to identify and relate two different geometric representations of the saddle points.

%%%%%%%%%%%%%%%
\subsection{dS representation of saddle points}
\label{sechomiso}
%%%%%%%%%%%%%%%

At large volume the wave function predicts an ensemble of classical, Lorentzian histories. Classical histories correspond to curves in the complex $\tau$-plane along which both the scale factor and field are real. It was found in \cite{Hartle2008} that in the above parameterisation all curves corresponding to classical histories asymptote to a constant value $z_f$\footnote{In \cite{Hartle2008} these curves were found by starting at the SP with a complex value of the scalar field $\phi(\tau_0)$. By tuning the phase of $\phi(\tau_0)$ together with the value of $z_f$ they found all asymptotically vertical curves $z=z_f$  along which $a$ and $\phi$ are both real and the classicality conditions satisfied. There is one such curve, defining one classical Lorentzian history, for each value $|\phi(\tau_0)| \equiv \phi_0$. In a single field potential this yields a one-parameter family of homogeneous and isotropic, asymptotically de Sitter, classical Lorentzian histories.}.

This can also be seen from the asymptotic expansions \eqref{aa_series} and \eqref{pphia}, which to leading order in $u$ can be written as
\be
a(u) = \frac{c}{u}= |c|e^{i\theta_c} e^{iHz}e^{Hy}, \qquad \phi(u) = \alpha u^{\lambda_{-}^{\ }} = |\alpha|e^{i\theta_{\alpha}}e^{-i\lambda_{-}^{\ }Hz}e^{-\lambda_{-}^{\ }Hy},
\label{lead_phase}
\ee
where $c$ and $\alpha$ are complex constants that are not determined by the asymptotic equations.
The condition that $a$ and $\phi$ are both real in the large volume limit along a constant value $z_f$ requires that
\be
\label{phases}
\theta_c=- H z_f, \qquad \theta_{\alpha}= \lambda_{-}^{\ } H z_f .
\ee
Then along the $z= z_f$ curve we have
\begin{equation}
ds^2 \approx - dy^2 + |c|^2  e^{2Hy} \gamma_{ij}dx^idx^j, \qquad \phi \approx \alpha e^{-\lambda_-^{\ } H y}.
\end{equation}
Hence along this contour $y$ acts as a time coordinate and the metric represents an asymptotic Lorentzian de Sitter universe with a slowly decaying scalar field.
The asymptotic contribution to the saddle point action is given by the integral \eqref{eucact_alt} along the curve  $z=z_f$. It is immediate that there will be no contribution to the amplitude $A$ of the wave function from this part of the contour: The integrand  in \eqref{eucact_alt} is real as is $d\tau=dy$. Instead this part of the contour yields a large negative contribution to the phase of the wave function, as required by the outgoing tunneling boundary conditions. Thus $\Psi_T$ oscillates rapidly with an approximately constant amplitude and describes an expanding, inflationary history.

\begin{figure}
\begin{center}
\includegraphics[width=8.5cm]{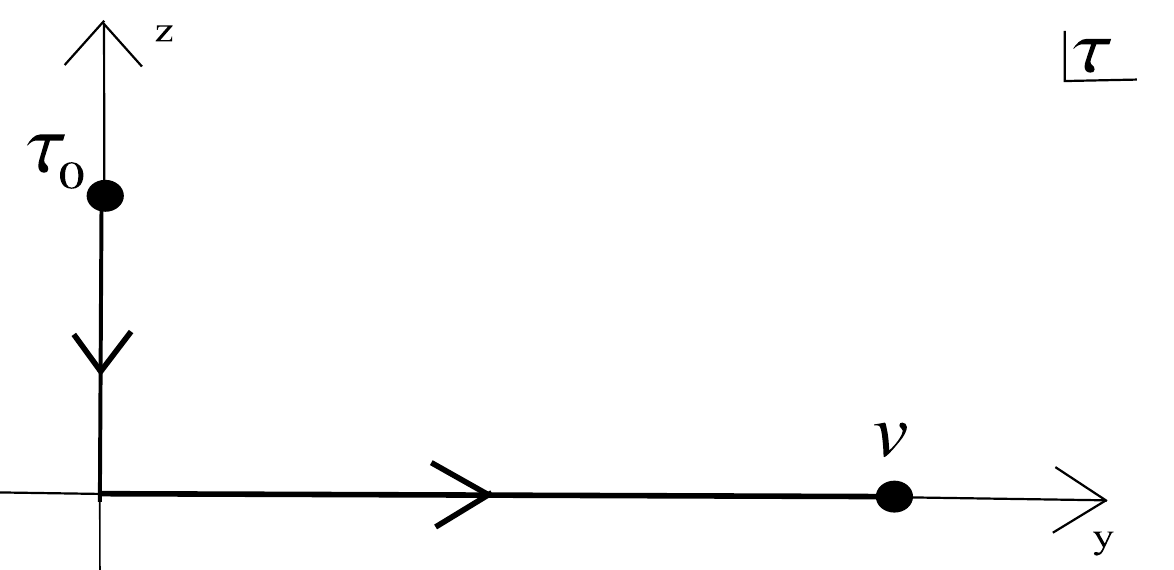} 
\includegraphics[width=8.5cm]{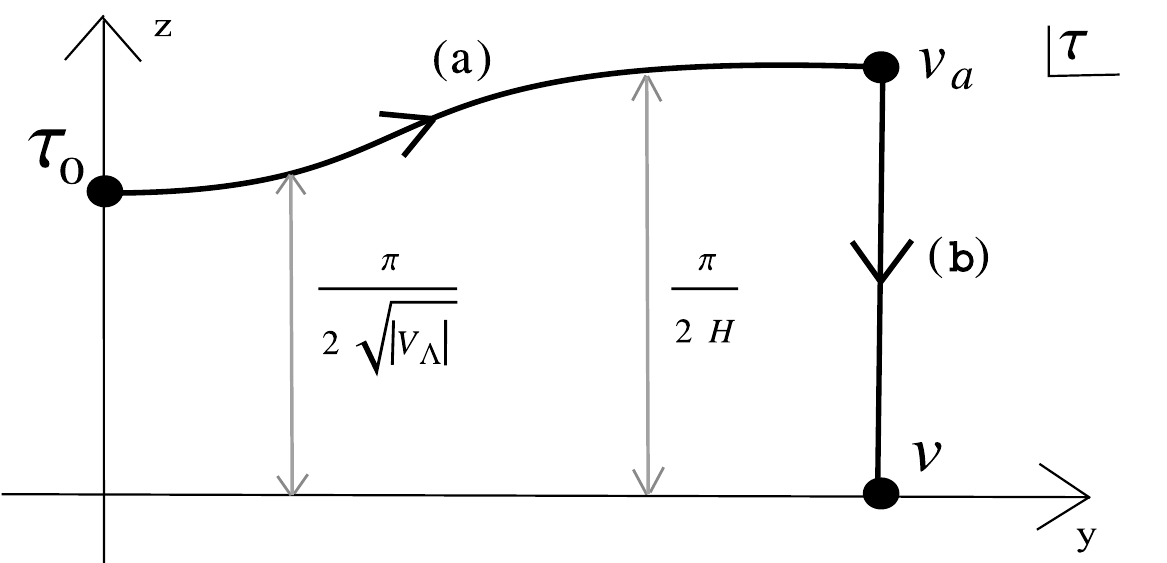} 
\end{center}
\caption{The dS contour $C_D$ (left) and the AdS contour $C_A$ (right) in the complex $\tau$-plane yield two distinct geometric representations of the same complex saddle point.}
\label{contours}
\end{figure}

It does not follow from the above analysis that the tuning \eqref{phases} is possible with regularity conditions at the origin and therefore that the saddle points actually exist. However \cite{Hartle2008} have shown they do. In the neighbourhood of the SP at $\tau = \tau_0$ the $O(4)$ invariant saddle point solutions take the form \begin{equation}
\phi(\tau) \approx \phi_0, \qquad a(\tau) \approx  \frac{\coth\left[\sqrt{V_\Lambda(\phi_0)} \tau\right]}{\sqrt{V_\Lambda (\phi_0)}}.
\label{norollsol}
\end{equation}
Taking $\tau_0 \approx i \pi /(2\sqrt{V_\Lambda (\phi_0)})$ and defining a contour $C_D$ that first runs from $\tau_0$ to $z_f=0$ along the z-axis and then along the $y$-axis yields a geometric representation of the saddle points in which an approximately Euclidean four sphere is smoothly joined onto a classical, expanding Lorentzian dS universe. This dS representation is illustrated in Fig \ref{contours}(a). 

The action integral over the Euclidean regime determines the amplitude $A$ of the corresponding classical history and is approximately given by \cite{Vilenkin1987} 
\begin{equation} \label{Ract}
\log A \approx - \frac{3\pi}{2 V_\Lambda (\phi_0)}.
\end{equation}
We can only write this explicitly in terms of the mini superspace coordinates, when an analytic solution is known along the entire contour. For example when the scalar field is relatively small at the SP and moving in a quadratic potential we have the analytic solution \cite{Hartle2008},
\begin{equation}
\phi = \chi \frac{_2 F_1[\lambda_-^{\ },\lambda_+^{\ },2,(1 + i \sinh H\tau)/2]}{_2 F_1[\lambda_-^{\ },\lambda_+^{\ },2,(1 + i \sinh [ \cosh^{-1}(H b)])/2]},
\end{equation}
%\begin{equation}
%\phi = \chi \frac{i + \sinh[ \cosh^{-1}(H b)]}{i + \sinh(H\tau)}.
%\end{equation}
where $ _2F_1$ is the hypergeometric function. This specifies a relation $\phi_0 = C \chi b^{\lambda_-^{\ }}$, where C is a constant which depends on $H,\lambda_-^{\ }$ and $\lambda_+{\ }$. In this case the amplitude of the wave function is given by
\begin{equation}
\log A \simeq - \frac{3\pi}{2H^2} + \frac{3\pi m^2 |C|^2 \chi^2 b^{2\lambda_-^{\ }}}{2 H^4} +\mathcal{O}(\chi^4).
\label{m2act}
\end{equation}

%%%%%%%%%%%%%%%%
%%%%%%%%%%%%%%%%

%%%%%%%%%%%%%%%%%%%%%%
\subsection{AdS representation of saddle points}
%%%%%%%%%%%%%%%%%%%%%%
\label{homoAdS}

The contour $C_D$ is not the only useful representation of the saddle points. The saddle point action is given by the integral \eqref{eucact_alt} and can be evaluated along any contour $C(\tau_0,\upsilon)$ connecting the SP to the endpoint $\upsilon$. Consider now the contour $C_A$ shown in Fig. \ref{contours}(b). In the neighbourhood of the SP the contour lies along 
\begin{equation}
z_{a}(y) \simeq  \frac{\pi}{2\sqrt{|V_\Lambda(\phi(y))|}}.
\label{zAdS}
\end{equation}
Hence the first branch (a) gradually moves away from the dS contour as the scalar field rolls down the hill. For large values of $y$ the contour asymptotes to $z_{a} = \pi/(2H)$. At the point $\upsilon_a \equiv i \pi/ (2H) + y_\upsilon$ it turns and runs vertically down along (b) towards the endpoint $\upsilon$. This contour has the same endpoint $\upsilon$, the same action, and makes the same predictions as $C_D$, but the saddle point geometry is different. Eq.\eqref{lead_phase} shows that the displacement from the real axis to $z_a$ replaces $u$ by $-iu$ and therefore $a(u)$ by $ia(u)$, to leading order.
Since $a$ was real along the real axis it will be imaginary along $z=z_a$. The asymptotic form of the metric \eqref{himet} along $z=z_a$ is
\begin{equation}
ds^2 \approx - dy^2 - |c|^2  e^{2Hy} \gamma_{ij}dx^i dx^j,
\end{equation}
and the asymptotic form of the scalar field along the $z=z_a$ curve is given by
\begin{equation}
\phi (y) \approx |\alpha | e^{-i \lambda_- H \pi /2}e^{-\lambda_- H y} \equiv \tilde{\alpha} e^{-\lambda_- H y}. 
\label{phase}
\end{equation}
Hence the saddle point geometry along this part of the contour is that of an asymptotically AdS, spherically symmetric domain wall with a complex scalar field profile in the radial direction $y$. The negative signature means that along this part of $C_A$ the action \eqref{lact} acts as that of Einstein gravity coupled to a negative cosmological constant $-\Lambda$ and a negative potential $-V$, which explains why the AdS behaviour emerges. Eq. \eqref{phase} shows that the asymptotic phase of the scalar field along part (a) of the contour is universal and determined by the boundary condition that it is asymptotically real along the $z=0$ curve. 

\begin{figure}
\begin{center}
\includegraphics[width=8.5cm]{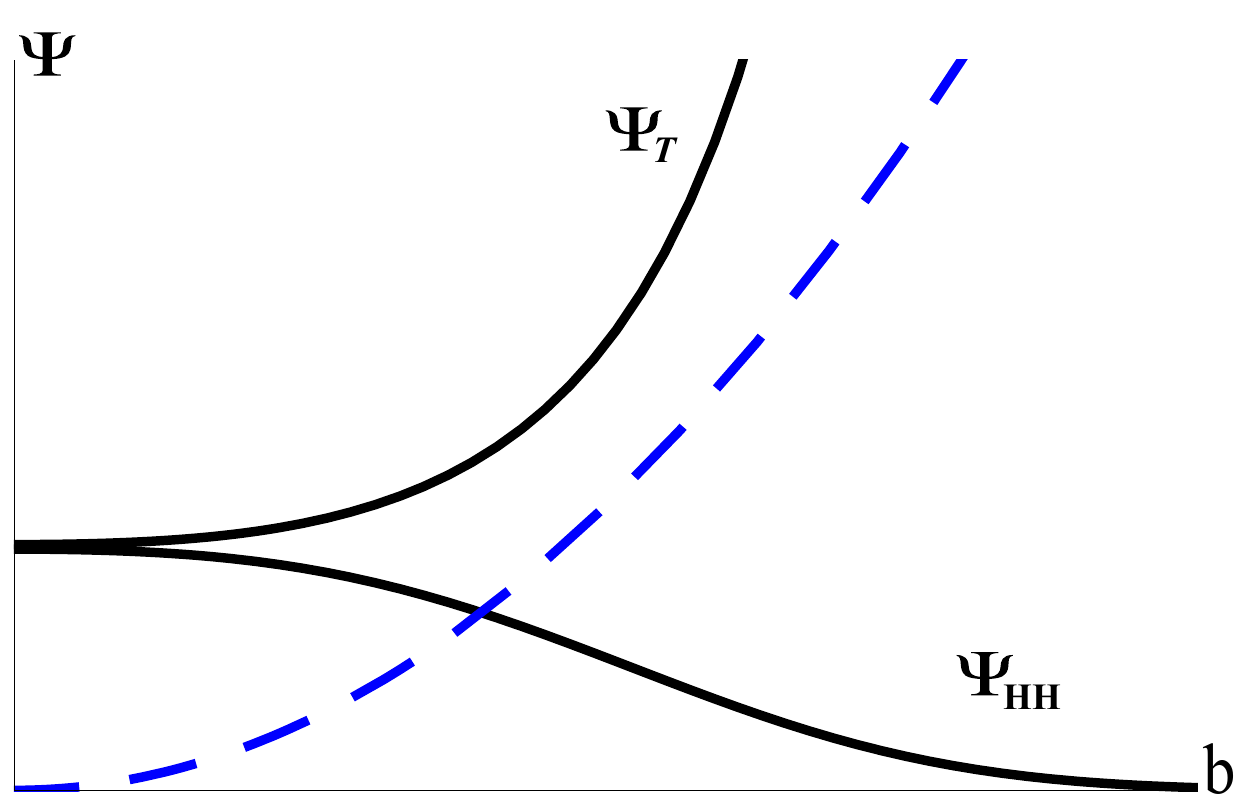} 
\end{center}
\caption{Qualitative behavior of $\Psi_T$ and $\Psi_{HH}$ in the AdS regime of the saddle points along branch $(a)$ of the contour $C_A$. The tunneling state corresponds to the usual growing wave function in AdS familiar from AdS/CFT whereas $\Psi_{HH}$ is a decaying wave function in its AdS regime. The blue dashed line shows the superpotential $U$.}
\label{wfads}
\end{figure}

The contribution $i S_{L(a)}$ to the saddle point action $iS_L$ coming from the integral along (a) is equal to minus the Euclidean AdS action of the domain wall solution and therefore exhibits the usual volume divergences associated with AdS. In this AdS domain the tunneling wave function thus behaves as the growing wave function familiar from AdS/CFT. This is in sharp contrast with the Hartle-Hawking wave function which behaves as a decaying wave function in its AdS domain \cite{Hertog2011}. We illustrate this difference in Fig.\ref{wfads}.
 
%%%%%%%%%%%%%%%%
%%%%%%%%%%%%%%%%

The contribution to the saddle point action from the vertical closing of the contour (b) regulates the divergences. This follows immediately from the fact that the amplitude of $\Psi_T$ along the horizontal part of the dS contour tends to a constant. The contribution from (b) therefore must cancel the divergences from (a), and also provide much of the phase of the wave function.

There remains the relation between the finite `regulated' action $iS_{L(a)}^{reg}= -I^{reg}_{DW}$ on (a), where  $I^{reg}_{DW}$ is the usual regulated Euclidean AdS action of the domain wall, and the amplitude $A$ of the wave function evaluated at the endpoint $\upsilon$. This connection is supplied by the action integral \eqref{eucact_alt} along (b).  In the next section we show that for all asymptotically locally de Sitter saddle points, i.e. including those corresponding to inhomogeneous final configurations,  the vertical branch (b) does not contribute to the amplitude.

%%%%%%%%%%%%%%%
\section{General Saddle points}
%%%%%%%%%%%%%%%
\label{asymptw}

The above discussion is not restricted to minisuperspace. It extends to saddle points corresponding to general boundary configurations specified by complex metrics of the form \eqref{sad}. In terms of the variable $u$ the large volume expansion of the general complex solution of the Einstein equation is given by \cite{Hertog2011}
\begin{align}
g_{ij}(u,\vec{x}) &= \frac{c^2}{u^2} \left[ \tilde{h}_{ij}^{(0)}(\vec{x})+ \tilde{h}_{ij}^{(2)}(\vec{x}) u^2  + \tilde{h}_{ij}^{(-)}(\vec{x}) u^{2\lambda_-}+ \cdots + \tilde{h}_{ij}^{(3)}(\vec{x}) u^{3} + \cdots \right], \label{a_series}\\
\phi(u,\vec{x}) &=  u^{\lambda _-}( \alpha(\vec{x}) + \alpha_1(\vec{x}) u + \cdots )   +  u^{\lambda _+}(\beta(\vec{x}) + \beta_1(\vec{x}) u + \cdots),
\label{phia}
\end{align}
where $\tilde{h}_{ij}^{(0)}(\vec{x})$ has unit volume.
As in the homogeneous case, the asymptotic solutions are specified by the asymptotic equations in terms of the boundary functions $c^2 \th_{ij}$ and $\alpha$, up to the $u^3$ term in \eqref{a_series} and to order $u^{\lambda_{+}^{\ }}$ in \eqref{phia}. Beyond this the interior dynamics and the boundary condition of regularity on $\mathcal{M}$ become important.
 
In saddle points associated with asymptotically classical histories the phases at the origin are tuned so that $g_{ij}$ and $\phi$ become real for small $u$ along the horizontal part of a contour $C_D$ at $z=z_f$. Since the expansions are analytic functions of $u$ that means there is again an alternative contour $C_A$ that asymptotically runs at $z_a=z_f+\pi/(2\sqrt{|V_\Lambda}|)$ along which the metric $g_{ij}$ is also real, but with the opposite signature. Thus we recover more generally the same story as in the homogeneous and isotropic example. 

Furthermore, it was shown in \cite{Hertog2011} that the asymptotically finite contribution to $i S_L$ coming from the first branch of $C_A$ is the same as the logarithm of the saddle point amplitude at the endpoint $\upsilon$. To see this one can evaluate the action integral along the vertical branch (b) of the contour connecting (a) to $\upsilon$, order by order in $u$. The on-shell action integral along the vertical part is given by
\be
\label{horizhij}
S_{L(b)}(\upsilon_a,\upsilon)=-\frac{1}{8\pi}\int_{z_a}^{z_f}dz\int d^3x \sqrt{-g} \left[6H^2 -{^3}R + 12V(\phi) +6({\vec\nabla}\phi)^2 \right],
\ee
where ${^3}R$ is the scalar three curvature of $g_{ij}$. As shown in \cite{Hertog2011} the asymptotic Einstein equation implies this does not contribute to the amplitude of the wave function in the large $y_\upsilon$ limit. This means $iS_{L(b)}$ only regulates the divergences of the action from (a) and supplies the phase necessary for classicality at $\upsilon$. In particular one has
\be
\label{horizhij2}
iS_{L(b)}(\upsilon_a,\upsilon) = (i-1)(I_1 +I_2+I_{\phi})(\upsilon_a)+{\cal O}(e^{- H y_\upsilon}),
\ee
where $I_1$ and $I_2$ are the familiar (real) gravitational counterterms and $I_{\phi}$ are additional (complex) scalar field counterterms which cancel the divergences arising from the slow fall-off of $\phi$ for large $y_\upsilon$ \cite{Hertog2011}. Hence for sufficiently large $y_\upsilon$ the combination 
\be
i S_{L(a)}(\upsilon_a) -(I_1+I_2+I_{\phi})(\upsilon_a)  \rightarrow -I^{\rm reg}_{DW}.
\label{reg}
\end{equation}
Summing the contributions from (a) and (b) yields 
\be
i S_L [b,\th_{ij}(\vx),\chi(\vx)] = -I^{\rm reg}_{DW} [\th_{ij}(\vx),\tilde{\alpha}(\vx)] + iS_{ct}[b, \th_{ij}(\vx),\alpha(\vx)] +  {\cal O}(e^{-Hy_\upsilon}),
\label{scT}
\ee
where $iS_{ct} \equiv (I_1+I_2+I_{\phi})(\upsilon)$. 

In this expression $\tilde{\alpha}$ is determined by the argument of the wave function as described in Eq.\eqref{phase}. Thus we find that in the limit of large scale factor the tunneling probabilities for all Lorentzian, asymptotically locally de Sitter histories with scalar matter are specified by the action of an ensemble of Euclidean AdS domain wall saddle points with a complex scalar field profile. This result leads directly to a holographic form of $\Psi_T$ as we discuss next.

%%%%%%%%%%%%%%%
\section{Holographic tunneling wave function}
%%%%%%%%%%%%%%%
\label{holo}

The AdS representation of the saddle points provides a natural connection between $\Psi_T$ in the large volume limit and Euclidean AdS/CFT. In the supergravity approximation the Euclidean AdS/CFT dictionary states that
\be
\exp (-I^{reg}_{DW}[\tilde h_{ij},\tilde{\alpha]}) = Z_{QFT}[\tilde h_{ij},\tilde{\alpha}] \equiv \langle \exp \int d^3x \sqrt{\tilde h} \tilde{\alpha} {\cal O} \rangle_{QFT},
\label{operator}
\ee
where the dual QFT lives on the conformal boundary represented here by the three-metric $\th_{ij}$. For radial domain walls this is the round three-sphere, but in general $\tilde{\alpha}$ and $\tilde h_{ij}$ are arbitrary functions of all boundary coordinates $\vx$. 

Applying \eqref{operator} to \eqref{scT} yields the following holographic formulation of the asymptotic tunneling wave function,
\beq
\Psi_T[h,\chi] = Z_{QFT}[\tilde{h},\tilde{\alpha}]e^{iS_{ct}[h,\chi]},
\label{holp}
\eeq
in which the probabilities of boundary configurations in the large volume limit are given by the partition function of CFTs defined on the AdS boundary and deformed by certain operators\footnote{It has been suggested that the wave function at finite scale factor can be obtained from an RG flow in the dual \cite{Strominger2001b}.}.

As discussed above the sources $(\tilde h, \tilde{\alpha})$ of $Z_{QFT}$ in \eqref{holp} are locally related to the argument $(\tilde h,\chi)$ of the wave function. The dependence of the field theory partition function on the sources gives a measure on different asymptotically locally de Sitter configurations. For sufficiently small values of the matter sources and sufficiently mild deformation of the round three sphere geometry one expects the integral defining the partition function to converge. 

%%%%%%%%%%%%%%%%%%%%%%%%%%%%%%%
Eq. \eqref{holp} is an example of a dS/CFT duality, albeit at the semiclassical level only. By this we mean that the derivation leading to \eqref{holp} is based on the saddle point approximation of the bulk wave function, and hence concerns the large $N$ limit of any dual field theory. It is an important open question whether a dS/CFT duality like \eqref{holp} holds when loop corrections in the bulk are taken into account.

A concrete dynamical model to which the duality \eqref{holp} applies is provided by ${\cal N}=8$ gauge supergravity in four dimensions, which admits an $AdS_4$ vacuum solution and is dual to ABJM theory in the large $N$ limit. The bulk theory contains negative mass scalars and admits several consistent truncations to AdS gravity coupled to one or more $m^2/H^2 =-2$ scalars. These scalars act as light, positive mass scalars in the dS regime of the bulk saddle points and thus can drive inflation in the corresponding Lorentzian history. The (complex) sources in the boundary theory that correspond to those scalars turn on finite, relevant deformations of the CFT. The duality \eqref{holp} states that the large N ABJM partition function as a function of those particular sources yields a dual way to compute the tunnelling measure in this model.

We should note, however, that the space of field theory deformations that are allowed is rather restricted. This is because one can only turn on those sources that preserve the asymptotic dS structure. This excludes in particular all irrelevant deformations corresponding to positive mass scalars in the AdS theory since these act as tachyonic scalars on the dS branch of the contour which, when turned on, destroy the asymptotic dS structure. Even scalars in dS with positive masses larger than $+9H^2/4$ are difficult to incorporate. This is because if they don't decay they can form stable bound halos, slowing or even reversing the expansion in local regions of the universe thereby again altering the asymptotic structure. In the AdS regime such fields behave as tachyonic scalars with masses below the Breitenlohner-Freedman bound. Any AdS theory with such scalars would admit bubble solutions that describe the decay of the AdS vacuum, consistent with the absence of a well-defined asymptotic dS structure.
%%%%%%%%%%%%%%%%%%%%%%%%%%%%%%%%%%%%

%A concrete example of the semiclassical dS/CFT duality \eqref{holp} is provided by ${\cal N}=8$ gauge supergravity in four dimensions, which admits an $AdS_4$ vacuum solution and is dual to ABJM theory. The bulk theory contains negative mass scalars and admits several consistent truncations to AdS gravity coupled to one or more $m^2/H^2 =-2$ scalars. The corresponding sources in the boundary theory turn on finite, relevant deformations of the CFT. These act as light, positive mass scalars in the dS regime of the saddle points that can drive inflation in the corresponding Lorentzian history. 
%
%Evidently in dS/CFT one can only turn on those sources that preserve the asymptotic dS structure. This is a severe restriction which excludes in particular all irrelevant deformations corresponding to positive mass scalars in the AdS theory. These would act as tachyonic scalars on the dS branch of the contour which, when turned on, destroy the asymptotic dS structure. Even scalars in dS with positive masses larger than $+9H^2/4$ are difficult to incorporate. This is because if they don't decay they can form stable bound halos, slowing or even reversing the expansion in local regions of the universe thereby again altering the asymptotic structure. In the AdS regime such fields behave as tachyonic scalars with masses below the Breitenlohner-Freedman bound. Any AdS theory with such scalars would admit bubble solutions that describe the decay of the AdS vacuum, consistent with the absence of a well-defined asymptotic dS structure.

In summary, the asymptotic dS structure which is a prerequisite in any dS/CFT proposal acts as a final condition that strongly constraints the possible deformations and therefore the dynamics.

%%%%%%%%%%%%%%%
\section{Testing the duality}
%%%%%%%%%%%%%%%
\label{duals}

It is not feasible at present to compute the partition function for deformed CFTs on $S^3$ that are dual to Einstein gravity. To gain intuition and support for the duality \eqref{holp} we therefore consider the simpler Klebanov-Polyakov version of the correspondence wherein the large $N$ field theory is tractable \cite{Klebanov2002}. This conjectures a duality between a higher spin gravity in $AdS_4$ and the singlet sector of the critical $O(N)$ vector model at large $N$ in three dimensions. In the spirit of \cite{Hartnoll2005,Anninos2012} we will compare exact results on the field theory side with properties of the bulk wave function in Einstein gravity, and interpret a qualitatively similar behavior as a positive test of the above conjectured duality.

Specifically we consider the partition function of small but finite, constant mass deformation of the critical $O(N)$ vector model on $S^3$. Since the operator that is sourced has dimension two we take the bulk to be four dimensional Einstein gravity theory coupled to a positive cosmological constant and a scalar field of mass $m^2= 2H^2$. Along the AdS branch of the saddle points the asymptotic expansion of the scalar field is then 
\begin{equation}
\phi = \tilde{\alpha} e^{-Hy} +\tilde{\beta}e^{-2 H y},
\end{equation}
where $\tilde{\alpha}=-i\alpha$ and $\tilde{\beta}=-\beta$. We will compare the partition function as a function of a mass deformation of the critical $O(N)$ model sourced by $\tilde{\alpha}$ with the asymptotic bulk wave function in the field basis as a function of the scalar boundary value $\alpha /(2 H b)$.

Before doing so, however, we first consider a mass deformation of the simpler, minimal $O(N)$ vector model. The partition function of this is related to that of the critical model by a double trace RG flow \cite{Klebanov2011, Anninos2011}. In the minimal model a mass deformation is sourced by $\tilde{\beta}$ multiplying an operator of conformal dimension one. In the bulk this means one adopts the alternative quantisation in which $\tilde \beta$ is the source, corresponding to calculating the bulk wave function in a different basis. Given the subtleties with the signs of the coefficients in the complex saddle points it is instructive to illustrate how the duality works in both bases, and thus to consider both the critical and the minimal model.

\subsection{Minimal O(N) vector model}
First we consider the action of the minimal $O(N)$ vector model on a three sphere of radius $r$ and a finite constant mass deformation $\mu^2$, given by \cite{Klebanov2011}
\beq
S=\frac{1}{2} \int d^3x \sqrt{g} \left[ (\nabla \vec{\Phi})^2 +\frac{R}{8} \vec{\Phi} ^2 + \mu^2\vec{\Phi}^2 \right]\,,
\label{cftact}
\eeq  
where $\vec{\Phi}$ is an $N$-component field which transforms as a vector under $O(N)$ rotations,  $R$ is the Ricci scalar and $\nabla$ is the covariant Laplacian operator of $S^3$. 

The free energy of this model is
\beq \label{freeEdiv}
-F=\log Z_{min}[\mu^2]=-\frac{N}{2}\log \det \left[\frac{-\Box+\frac{R}{8}+\mu^2}{\epsilon^2}\right]
\eeq
where $\epsilon$ is an arbitrary dimensionful constant, interpreted as a sliding renormalization scale. 
The functional determinant can be calculated using a zeta function regularization scheme. This yields \cite{Klebanov2011}
\beq \label{freeE} 
\begin{split}
\log Z_{min}[\mu^2]=-\frac{N}{48 \pi^2} \Big\{ 6\pi ^2 (1-4r^2 \mu^2)\log\lp 1-e^{-i\pi\sqrt{1-4 r^2 \mu^2}}\rp +12 Li_3\lp e^{-i\pi\sqrt{1-4r^2 \mu^2}}\rp+\\ i\pi\sqrt{1-4r^2\mu^2} \left[ \pi^2 (1-4r^2\mu^2 )+12 Li_2\lp e^{-i\pi\sqrt{1-4r^2\mu^2}}\rp\right]\Big\}.
\end{split}
\eeq
For small $r^2\mu^2$ this becomes
\beq
\log Z_{min}[\mu^2]\sim - \frac{N}{8}\left[ \log 4-\frac{3 \zeta(3)}{\pi ^2}\right] + \frac{N}{16}  r^4 \pi^2 \mu^4  + \mathcal{O}(r^6 \mu^6).
 \label{Ztocomp}
\eeq
The first term is the partition function of the massless $O(N)$ vector model on $S^3$. The second term is the change in the partition function induced by a small but finite mass term that deforms the theory away from its $r^2 \mu^2 =0$ fixed point. 

To show that $Z_{min}$ exhibits the same behaviour as $\Psi_T$ we first perform a canonical transformation to write $\Psi_T$ as a function of $\beta$. It suffices for our purposes here to consider the small scalar field regime described at the end of Section \ref{sechomiso}. We write the asymptotic scalar profile as
\begin{equation}
\phi = \frac{\alpha}{2H} a^{-1} + \frac{\beta}{4H^2} a^{-2} \equiv \hat{\alpha} a^{-1} + \hat{\beta} a^{-2}.
\label{phiexp}
\end{equation}
The general form of a canonical transformation at the level of the action is
\begin{equation}
\Pi d\phi - \mathcal{H}(\phi,\Pi,\lambda) =  B d\hat{\beta} - \mathcal{K}(\hat{\beta},B,\lambda) + \frac{dG(\phi,\hat{\beta)}}{dt},
\end{equation}
where $\Pi = \dot{\phi}a^3$ is the conjugate momentum of $\phi$, and $B = \hat{\alpha}H$  the conjugate momentum of $\hat{\beta}$. Also, $\mathcal{H}$ and $\mathcal{K}$ are the respective Hamiltonians and $G$ is the generating function. This formulation is derived by requiring that the transformed action gives the same equations of motion. That is, the variation of both actions is the same up to a total derivative. 

We are interested in a canonical transformation of the form
\begin{equation}
\left(\begin{matrix}
\phi \\ \Pi
\end{matrix}\right)
=
\left(\begin{matrix}
a^{-2} && { H^{-1}}a^{-1} \\ -2 H a && -a^2
\end{matrix}\right)
\left(\begin{matrix}
\hat{\beta} \\ B
\end{matrix}\right).
\end{equation}
To find the generating function we evaluate
\begin{align}
\int \Pi d\phi 
&= \int \left[\left(3 { H^2} a^{-1} \hat{\beta}^2 +4 {H} \hat{\beta} B + \frac{3}{2}a B^2\right)d\lambda + B d\hat{\beta}\right] + \left[ - H a^{-1} \hat{\beta}^2 - 2 \hat{\beta} B - \frac{1}{2} { H^{-1}} a B^2 \right]_{BD}.
\end{align}
The boundary term is the generating function $G$, when expressed as a function of $\hat{\beta}$ and $\chi$ only. The wave functions in different bases are thus related by
\begin{align}
\Psi_T(\hat{\beta},b) &= \int  d \chi \exp\left[\frac{3\pi i{ H}}{4}(\chi^2 b^3 + 2\hat{\beta} \chi b - \hat{\beta}^2 b^{-1})\right]  \Psi_T(\chi,b).
\label{transwf}
\end{align}
We showed in section \ref{sechomiso} that the scalar part of $\Psi_T(\chi,b)$ is given by\footnote{Here we use that $C=1/(2iH)$ for $m^2 = 2 H^2$, with $C$ defined above eq.\eqref{m2act}.}
\begin{equation}
\Psi_T(\chi,b)= \exp\left[\frac{3\pi}{4} \left(-i H \chi^2 b^3 + \chi^2 b^2 \right) \right].
\end{equation}
Hence we find
\begin{equation}
\Psi_T(\hat{\beta},b) = \int  d \chi \exp \left[\frac{3\pi }{4}(\chi^2 b^2  +  2i H\hat{\beta} \chi b + \mathcal{O}( b^{-1}))\right]  .
\end{equation}
A steepest descent approximation of the integral yields the relation $\chi = - i H\hat{\beta} b^{-1}$. The amplitude of $\Psi_T$ in terms of $\beta$ is thus given by
\begin{equation}
\log A[\beta] = -\frac{3\pi}{2H^2}+\frac{3 \pi \beta^2}{64 H^2} + \mathcal{O}(\beta^4).
\end{equation}
Remarkably this is qualitatively similar to the behavior of $Z_{min}$ in \eqref{Ztocomp}, since the holographic dictionary relates $N \sim H^{-2}$  and $\mu^2 \sim \tilde{\beta} = -\beta$. We conclude that $\log Z_{min}[\mu^2] \simeq \log A[\beta]$, in agreement with our general result \eqref{holp}.

\subsection{Critical O(N) vector model}
We now proceed to compute the partition function of the critical $O(N)$ vector model. This model can be obtained from the free theory by deforming it with a relevant double trace deformation $\lambda (\vec{\Phi} \cdot \vec{\Phi})^2/(8N)$, where $
\lambda$ is a constant of conformal dimension one, and taking the dimensionless coupling $ r\lambda \rightarrow \infty$. We include an additional single-trace deformation $\vec{\Phi} \cdot \vec{\Phi}$ with coefficient $\lambda \sigma$, where $\sigma$ corresponds to the asymptotic bulk coefficient $\tilde \alpha$. The action is \cite{Klebanov2002, Hartnoll2005}
\beq
S= \frac{1}{2} \int d^3x \sqrt{g} \left[  (\nabla \vec{\Phi})^2 +\left(\frac{1}{8}R +\lambda \sigma\right) \vec{\Phi} ^2 + \frac{\lambda (\vec{\Phi} \cdot \vec{\Phi} )^2}{2 N} \right]\,.
\label{crcftact}
\eeq 
In the limit $r\lambda \rightarrow \infty$ the partition function of this model is related to the partition function of the minimal model by a basis transformation \cite{Anninos2012}, similar to the transformation of $\Psi_T$ above. To compute this one introduces an auxiliary field $N \tilde{\mu}=\vec{\Phi}\cdot\vec{\Phi}$, such that the action can be rewritten in terms of single trace operators
\beq
S= \frac{1}{2} \int d^3x \sqrt{g} \left[  (\nabla \vec{\Phi})^2 +\left(\frac{1}{8}R+\lambda \sigma+\lambda \tilde{\mu}\right) \vec{\Phi} ^2 - \frac{N\lambda \tilde{\mu}^2}{2} \right].
\label{crcftact2}
\eeq 
Integrating out the $\vec{\Phi}$ field one obtains
\beq \label{pfunccrit}
Z_{crit}[\sigma] = e^{\frac{N \lambda}{2}\int d\Omega_3 \sigma^2} \int \mathcal{D}\mu^2 \exp\left[ N \int d\Omega_3 \lp \frac{\mu^4}{2\lambda}-\sigma \mu^2 \rp \right] Z_{min}[\mu^2]\,,
\eeq
where $\mu^2 \equiv \lambda \sigma+\lambda \tilde{\mu}$. The equivalence with Eq.\eqref{transwf} shows that both models are indeed related by a Fourier type transformation. The first factor on the r.h.s. in \eqref{pfunccrit} is local. As we discussed this is canceled by the contribution to the action integral along the second, vertical part of the saddle point contour. At large $N$ one can evaluate the integral \eqref{pfunccrit} in the saddle point approximation. Using \eqref{freeE} the saddle point equation reads
\beq \label{speq}
\frac{16\pi r^2\mu^2}{r\lambda}+16 \pi r \sigma  = \sqrt{1-4r^2\mu^2} \cot{\lp \frac{\pi}{2}\sqrt{1-4r^2\mu^2}\rp}\,.
\eeq
In the large $r \lambda$ regime the first term vanishes and the saddle point equation can be solved analytically. For small dimensionless deformations $r \sigma \ll 1$  eq.\eqref{speq} implies that $r^2\mu^2 \ll 1$, which then yields 
\beq\label{Ztocompcr}
\log{Z_{crit}[\sigma]} \sim - \frac{N}{8}\left[ \log 4-\frac{3 \zeta(3)}{\pi ^2}\right] - N  \pi^2 r^2 \sigma ^2 + \mathcal{O}(r^3\sigma^3).
\eeq
On the other hand, the amplitude of $\Psi_T$ in the field basis in this regime is given by (cf. \eqref{m2act})
\beq
\log A[\alpha] = -\frac{3\pi}{2H^2}+\frac{3 \pi \alpha^2 }{16H^2} + \mathcal{O}(\alpha^4),
\label{cIR}
\eeq 
where we used that $C= 1/(2iH)$ for $m^2 = 2 H^2$ and $\chi = \alpha/(2Hb)$. Since $N \sim H^{-2}$ and $\sigma \sim \tilde \alpha=-i\alpha$ we again find qualitative agreement between the bulk and boundary calculations.

\section{Discussion}
\label{disc}

We have shown that the complex structure of the bulk saddle points specifying the semiclassical tunneling wave function in cosmology allows one to use Euclidean AdS/CFT to derive a dual formulation of $\Psi_T$. In this, the relative probabilities of asymptotically locally de Sitter configurations in the tunneling state are given in terms of the partition function of AdS/CFT duals defined on the conformal boundary and deformed by certain relevant operators.

Our derivation applies to general, inhomogeneous boundary configurations. It is therefore legitimate and interesting to ask whether the holographic form of $\Psi_T$ predicts fluctuations away from homogeneity are damped as required by observation. It is immediately clear from \eqref{holp} that this is not the case, because the two point functions of the AdS/CFT dual operators are positive \cite{Maldacena2002}. Hence holography indicates that the tunneling state does not yield a well-defined, normalisable cosmological measure beyond minisuperspace.

Independently of any application to cosmology our analysis shows that in the WKB approximation, $\Psi_T$ and $\Psi_{HH}$ can also be viewed as wave functions with asymptotic AdS boundary conditions. This is because their complex saddle points have a geometric representation in which their interior geometry is locally AdS. The tunneling wave function corresponds to the usual growing wave function in its AdS domain that features in AdS/CFT applications. By contrast the Hartle-Hawking wave function is a decaying wave function in its AdS domain. At the semiclassical level the probability distributions they predict are inversely related to each other. Indeed in \cite{Hertog2011} the Hartle-Hawking measure involves the inverse of the AdS/CFT dual partition function. But this simple symmetry is unlikely to hold beyond tree level; there is no reason why the loop corrections to both wave functions should be inversely related to each other. Instead one would expect that a complete dS/CFT framework for Einstein gravity that is rooted in Euclidean AdS/CFT will require a direct understanding of the decaying branch of the bulk wave function in AdS/CFT.

Having said this, a somewhat similar inverse relation shows up in the higher spin realisation of dS/CFT, where the partition function of the $Sp(N)$ model as a function of certain deformations (often) is the inverse of the original $O(N)$ partition function. Yet recent calculations of finite $N$ partition functions in this context do not conclusively settle whether the field theory describes $\Psi_{HH}$, $\Psi_T$ {\rx or} yet another state. In fact they hint at the possibility that different choices of boundary conditions on the fermions in the dual may provide the freedom needed to model different bulk wave functions. Our results clarify at least the bulk side of this question.

\vskip .2in

\noindent{\bf Acknowledgments:} We thank Dionysios Anninos, Nikolay Bobev, Frederik Denef, Daniel Harlow, James Hartle, Juan Maldacena and Yannick Vreys for discussions. TH and EvdW thank the KITP and the Physics Department at UCSB for their hospitality. This work is supported in part by the National Science Foundation of Belgium (FWO) grant G.001.12 Odysseus. TH is supported by the European Research Council grant no. ERC-2013-CoG 616732 HoloQosmos. We also acknowledge support from the Belgian Federal Science Policy Office through the Inter-University Attraction Pole P7/37 and from the European Science
Foundation through the  `Holograv' Network.

\end{document}